\documentclass[conference]{IEEEtran}
\IEEEoverridecommandlockouts
\usepackage{epsfig}
\usepackage{epsf}
\usepackage{psfrag}
\usepackage{cite}
\def\blfootnote{\xdef\@thefnmark{}\@footnotetext}
\usepackage[cmex10]{amsmath}

\usepackage{stfloats}
\begin{document}

\title{Coding Strategies for Noise-Free Relay Cascades with Half-Duplex Constraint}

\author{\IEEEauthorblockN{Tobias Lutz, Christoph Hausl, and Ralf K\"otter}
\IEEEauthorblockA{Institute for Communications Engineering, TU M\"unchen, Munich, Germany}
Email: \{tobi.lutz, christoph.hausl, ralf.koetter\}@tum.de
\thanks{This work was supported by the European Commission in the framework of the FP7 (contract n. 215252) and by DARPA under the ITMANET program.}}
\maketitle

\begin{abstract}
Two types of noise-free relay cascades are investigated. Networks where a source communicates with a distant receiver via a cascade of half-duplex constrained relays, and networks where not only the source but also a single relay node intends to transmit information to the same destination. We introduce two relay channel models, capturing the half-duplex constraint, and within the framework of these models capacity is determined for the first network type. It turns out that capacity is significantly higher than the rates which are achievable with a straightforward time-sharing approach. A capacity achieving coding strategy is presented based on allocating the transmit and receive time slots of a node in dependence of the node's previously received data. For the networks of the second type, an upper bound to the rate region is derived from the cut-set bound. Further, achievability of the cut-set bound in the single relay case is shown given that the source rate exceeds a certain minimum value.
\end{abstract}

\IEEEpeerreviewmaketitle

\section{Introduction}
The focus of this paper is on half-duplex constrained relay line networks, i.~e. on multi-hopping networks where the intermediate relay nodes are arranged in a cascade and, further, are not able to transmit and receive simultaneously. We consider networks with a single source-destination pair and networks where in addition to the source a single relay node intends to transmit own information. Since the main interest is to gain a better understanding of half-duplex constrained transmission, we assume noiseless network links in order to avoid detraction from the actual topic.

The classical relay channel goes back to van der Meulen~\cite{meu71}. Further significant results concerning capacity and coding schemes were obtained in \cite{cov79}. More recently, the focus of attention shifted towards relay networks and an achievable rate formula for relay line networks with a single source-destination pair together with a random coding scheme appeared in \cite{XieKum05}. A comprehensive literature survey as well as a classification of random coding strategies is given in \cite{KraGasGup05}. There has also been work on determining the capacity or rate region of various half-duplex constrained relay channels \cite{mad05},\cite{kho03} and networks \cite{tou03}, however, under the assumption that the time-division schedule is determined a priori.

An obvious approach in order to handle the half-duplex constraint in a line network is to use a transmission scheme in which even numbered relays send in say even numbered time slots and receive during odd numbered time slots while odd numbered relays behave vice versa. If the source uses a binary alphabet, the rate becomes $0.5$ bits per use while a ternary alphabet yields a rate of $0.5 \hspace{1mm}\textrm{log}_2 3$ bits per use. By allowing randomly allocated transmit and receive time slots, higher rates are possible as was pointed out in \cite{kra04}. In~\cite{kra07}, the same author uses an entirely binary, noiseless model for the single relay channel such that the half-duplex constraint is included. It is shown there that capacity is equal to $0.7729$ bits per use what demonstrates that time-sharing falls considerably short of the theoretical achievability. By the way, the same channel model was used in \cite{van92} in a different context. Two coding schemes for this particular model were outlined therein, which, in hindsight, can be interpreted as half-duplex schemes.

We will introduce two further channel models for half-duplex constrained relays. Within the framework of these models, it is shown that the capacity of a half-duplex constrained single relay channel is equal to 0.8295 bits per use if the relay is able to distinguish binary symbols and 1.1389 bits per use if, in addition, the relay is capable of detecting time slots without transmission. Furthermore, it is shown that the capacity of each relay cascade with finite length is greater than one bit per use assumed the latter relay model is utilized. The key idea of the achievable scheme is to determine the slot allocation of each relay node in dependence of the data received by the relay before. With regard to half-duplex constrained line networks, where not only the source but also a single relay node intends to transmit own information to the same destination, an upper bound to the rate region is derived. We finally show that in the special case of a single relay channel (with source and relay source), a slightly different version of the introduced coding scheme is able to achieve a segment on the cut-set bound, provided that the source rate exceeds a certain minimum value.

\textit{Notation:} $|S|$ denotes the cardinality of set $S$ and $\mathcal{P}(S)$ the power set of $S$. Further, $S_{\bar{i}}:=S\backslash\{i\}$ while $\{f(i):1\leq i \leq m\}$ means $\{f(1),f(2),\dots,f(m)\}$. The conditional pmf $p_{Y|X}(y,x)$ is indicated as $p(y|x)$ whenever the random variables can be figured out from the arguments. Further, the vector $\mathbf{x}_{[0:m]}:=(x_0,x_1,\dots,x_m)$ summarizes realizations of the random variables $X_0,X_1,\dots,X_m$. The entropy expression $H(Y_i|X_{(k:k>1)})$ equals $H(Y_i|X_k)$ in case $k>1$ and $H(Y_i)$ in case $k \leq 1$. We will abbreviate $p_{X_i X_{i+1}}(a,b)$ as $p_{ab}^i$. 

\section{Network Model}
\begin{figure}[t]
\psfrag{X0}{\small{\textrm{$X_0$}}}
\psfrag{Xi}{\small{\textrm{$X_i$}}}
\psfrag{Xi-1}{\small{\textrm{$X_{i-1}$}}}
\psfrag{Yi}{\small{\textrm{$Y_i$}}}
\psfrag{Yi-1}{\small{\textrm{$Y_{i-1}$}}}
\psfrag{Y_m+1}{\small{\textrm{$Y_{m+1}$}}}
\psfrag{Relay}{\small{\textrm{$\small{\textrm{Relay}}$}}}
\psfrag{i}{\small{\textrm{${i}$}}}
\psfrag{i-1}{\small{\textrm{${i-1}$}}}
\psfrag{1}{\scriptsize{$1$}}
\psfrag{2}{\scriptsize{$2$}}
\centering
\epsfig{file=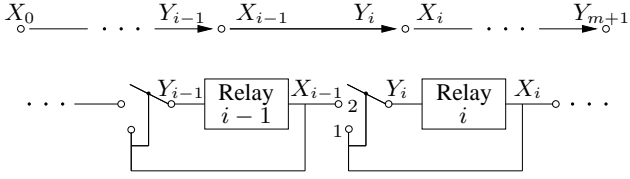, scale = 0.7}
\caption{The considered multiple relay cascade (top) and an excerpt. If relay~$i$ is transmitting, the switch is in position $1$ otherwise in position~$2$. }
\vspace{-0.5cm}
\label{fig:sys_model}
\end{figure}

We consider a discrete, memoryless line network composed of $m+2$ nodes whereas each node is characterized by a unique number from the integer set $\{0,\dots,m+1\}$. The integers $0$ and $m+1$ are allocated to source and destination, respectively. The remaining nodes $1$ to $m$ represent half-duplex constrained relays (abbreviated as HD relays). A graphical representation is given in Fig. \ref{fig:sys_model}. The output of the $i$th node, which is the input to the channel between node $i$ and $i+1$, is denoted as $X_i$, $i \in \{0,\dots,m\}$, and takes values on the alphabet $\mathcal{X}_i = \{0,1,\textrm{N}\}$, where N is meant to signify a channel use in which node $i$ is not transmitting. Correspondingly, the input of the $i$th node, which is the output of the channel between node $i-1$ and node $i$, is $Y_i$, $i \in \{1,\dots,m+1\}$, with values from the alphabet $\mathcal{Y}_i$. Each message $w_0$, sent via multiple hops from source node $0$ to sink node $m+1$, is uniformly drawn from the index set $\mathcal{W}_0 = \{1,\dots,2^{nR_0}\}$, where $n$ is the block length of the encoding scheme and $R_0$ the transmission rate. Apart from the source node, there is possibly a single relay node $r \in \{1,\cdots,m\}$, which intends to transmit independent indices taken from $\mathcal{W}_r = \{1,\dots,2^{nR_r}\}$ to the destination. Again, the transmission scheme is multi-hopping since the information flow associated with message $w_r$ has to pass all nodes with indices greater than $r$. We assume noiseless links what results in a deterministic network, i.~e. the entries in $p(\mathbf{y}_{[1:m+1]}|\mathbf{x}_{[0:m]})$ are either $0$ or $1$. 
In order to introduce the half-duplex constraint, we impose following channel model onto each relay node $i \in \{1,\dots,m\}$
\begin{equation}
Y_{i}=\left\{\begin{array}{ll} X_{i-1}, & \mbox{if }X_{i}=\textrm{N}\\ X_{i}, & \mbox{if }X_{i} \in \{0,1\}, \end{array}\right.
\label{relay_ch_model}
\end{equation}
where $Y_{m+1} = X_m$. Relay model (\ref{relay_ch_model}) is denoted as \textit{ternary} since the reception alphabet of each relay node is $\mathcal{Y}_{i} = \{0,1,\textrm{N}\}$. It can easily be verified that $\mathcal{Y}_{i} = \{0,1\}$ when $(\textrm{N,N})$ is excluded from the Cartesian product $\mathcal{X}_{i-1}\times\mathcal{X}_i$, and in this case the model is referred to as \textit{binary}. The interpretation of both models is as follows: in case relay $i$ sends binary data, i.~e. $x_{i} \in \{0,1\}$, it only hears itself and, thus, cannot listen to relay $i-1$ or, equivalently, relay $i$ and relay $i-1$ are disconnected. Conversely, if relay $i$ is quiet, i.~e. $x_{i} = \textrm{N}$, it is sensitive for the channel input of relay $i-1$. The feedback interpretation of the relay nodes as shown in Fig.~\ref{fig:sys_model} results from these considerations. As a consequence of the underlying model, the conditional channel pmf can be factored as 
\begin{eqnarray}
p\left(\mathbf{y}_{[1:m+1]}|\mathbf{x}_{[0:m]}\right) &=& p\left(y_1|\mathbf{x}_{[0:1]}\right)\cdots p\left(y_m|\mathbf{x}_{[m-1:m]}\right) \nonumber\\
&&p\left(y_{m+1}|x_m\right).
\end{eqnarray}
Moreover, we will assume that the channel inputs $X_0,X_1,\dots,X_m$ form a Markov chain what seems to be unmotivated at first glance but turns out to be without loss of optimality as explained in Remark \ref{Remark_3}.
\section{Coding Theorems}
\newtheorem{theorem}{Theorem}
\newtheorem{example}{Example}
\newtheorem{remark}{Remark}
\newtheorem{prop}{Proposition}
\begin{theorem}\label{th:C_one_source}
The zero-error capacity of the relay network defined above, where only the source but no relay transmits own information, is given by
\begin{equation}
C = \max_{p(\mathbf{x}_{[0:m]})} \min \left \{ H(Y_1|X_1), \dots, H(Y_m|X_m),H(Y_{m+1})\right\}.
\label{eq:C_one_source}
\end{equation}
\end{theorem}

\begin{proof}
The proof is given in the Appendix. Achievability is shown in the next section.
\end{proof}
\begin{example}[Single HD Relay Channel, $m=1$]\label{example_1}
The considered channel with a \textit{ternary} relay falls into the class of degraded relay channels~\cite{cov79}. At each time instance, the relay is either listening or transmitting. When the relay transmits, i.~e. $x_1 \in \{0,1\}$, the source input cannot be detected by the relay and, consequently, the source should not transmit. Thus, it can be assumed w.l.o.g. that $p_{00}^0 = p_{01}^0 = p_{10}^0 = p_{11}^0 = 0$. Hence, the source input is not random when $x_1 \in \{0,1\}$ and together with (\ref{relay_ch_model}), equation (\ref{eq:C_one_source}) reduces to
\begin{equation}
C = \max_{p(\mathbf{x}_{[0:1]})} \min \left \{H(X_0|X_1 = \textrm{N})p_{X_1}(\textrm{N}), H(X_1)\right\}.
\label{eq:C_single_relay_ex}
\end{equation}
However, when the relay is listening, i.~e. $x_1 = \textrm{N}$, the source should make optimum use of the channel by encoding with uniformly distributed input symbols, i.~e. $p_{0\scriptsize{\textrm{N}}}^0 = p_{1\scriptsize{\textrm{N}}}^0 = p_{\scriptsize{\textrm{N}\textrm{N}}}^0$. Furthermore, in order to achieve the maximum information flow $H(X_1)$ from the relay to the sink or, likewise, from a symmetry argument, we can choose $p_{\scriptsize{\textrm{N}}0}^0 = p_{\scriptsize{\textrm{N}}1}^0$. These considerations yield a single degree of freedom in (\ref{eq:C_single_relay_ex}). Since the maximum does not occur in the maximum of one of the two concave functions, (\ref{eq:C_single_relay_ex}) is solved by $H(X_0|X_1) = H(X_1)$. The resulting assignment is $p_{0\scriptsize{\textrm{N}}}^0 = p_{1\scriptsize{\textrm{N}}}^0 = p_{\scriptsize{\textrm{N}\textrm{N}}}^0 = 0.2395$ and $p_{\scriptsize{\textrm{N}}0}^0 = p_{\scriptsize{\textrm{N}}1}^0 = 0.1407$, which yields $C=1.1389$ bits per channel use.
\end{example}

\begin{remark}
Evaluation of capacity for the \textit{binary} HD model is almost along the same lines as in Example \ref{example_1}. However, the channel input $x_0x_1 = \textrm{NN}$ is not allowed in the \textit{binary} model and, thus, we a priori have $p_{\scriptsize{\textrm{N}\textrm{N}}}^0 = 0$, which yields $C = 0.8295$ bits per channel use.
\end{remark}
\begin{figure*}[!b]
\vspace{-0.3cm}
\small
\hrulefill
\setcounter{equation}{6}
\begin{equation}
R_0 + R_r \leq \max_{p(\mathbf{x}_{[0:m]})} \min \left\{\min \{H(Y_i|X_i):1 \leq i \leq r-1\} + \min \{H(Y_k|X_{(r-1:r-1\geq 1)},X_{(k:k\leq m)}):r+1 \leq k \leq m+1\}, H(Y_{m+1})\right\}
\label{sum_rate_two_sources}
\end{equation}
\end{figure*}
\begin{example}[Infinite HD Relay Channel, $m\rightarrow \infty$]\label{example_2}
All relays in the cascade behave according to the \textit{ternary} model. Due to the Markov property of the channel inputs, the joint pmf $p(\mathbf{x}_{[0:m]})$ is completely characterized by $p(\mathbf{x}_{[0:1]})$, $p(\mathbf{x}_{[1:2]})$, $\dots$, $p(\mathbf{x}_{[m-1:m]})$. Further, $H(Y_i|X_i) = H(X_{i-1}|X_i)$, which follows from (\ref{relay_ch_model}). The idea is now to find a probability assignment such that the $p(\mathbf{x}_{[i-1:i]})$ are equal for all $i \in \{1,2,\dots,m\}$ without violating any optimality requirements. If we can find such a probability assignment, capacity simply follows by maximizing a single $H(X_{i-1}|X_i)$ for that particular assignment. We now pick an arbitrary positive integer~$i$ and try to make $p_{kl}^{i-1}$ and $p_{kl}^{i}$ equal for all combinations $k,l \in \{0,1,\textrm{N}\}$. By the same arguments as in Example~\ref{example_1}, we can choose w.l.o.g. $p_{00}^{i-1} = p_{01}^{i-1} = p_{10}^{i-1} = p_{11}^{i-1} = 0$, and the same is valid for $p(\mathbf{x}_{[i:i+1]})$. As a simple consequence, $p_{\scriptsize{N}0}^{i-1} = p_{0\scriptsize{N}}^{i}$ and $p_{\scriptsize{N}1}^{i-1} = p_{1\scriptsize{N}}^{i}$ and, from a symmetry argument, $p_{\scriptsize{N}0}^{i-1} = p_{\scriptsize{N}1}^{i-1}$ and $p_{0\scriptsize{N}}^{i} = p_{1\scriptsize{N}}^{i}$. Further regarding our objective, we have to require that $p_{kN}^{i-1} = p_{kN}^{i}$ for $k \in \{0,1\}$. Since index~$i$ has been picked arbitrarily at the beginning, the procedure is valid for each  $p(\mathbf{x}_{[i-1:i]})$ and $p(\mathbf{x}_{[i:i+1]}), 1 \leq i \leq m-1$, what is sufficient in order to achieve equal pmfs with a common, single degree of freedom $(\textrm{e.~g. }p_{0\scriptsize{\textrm{N}}}^i)$. Hence, $H(X_{i-1}|X_i)$, $1 \leq i \leq m$, is easy to optimize yielding $H(X_{i-1}|X_i) = 1$~bit achieved at $p_{0\scriptsize{\textrm{N}}}^i = \frac{1}{6}$. The capacity $C$ is, therefore, equal to $1$~bit per channel use.
\end{example}

\begin{figure}[t]
\psfrag{(a)}{\footnotesize{(a)}}
\psfrag{(b)}{\footnotesize{(b)}}
\psfrag{(c)}{\footnotesize{(c)}}
\psfrag{R0}{\footnotesize{\textrm{$R_0$ (bits per use)}}}
\psfrag{R1}{\footnotesize{\textrm{$R_1$ (bits per use)}}}
\centering
\epsfig{file=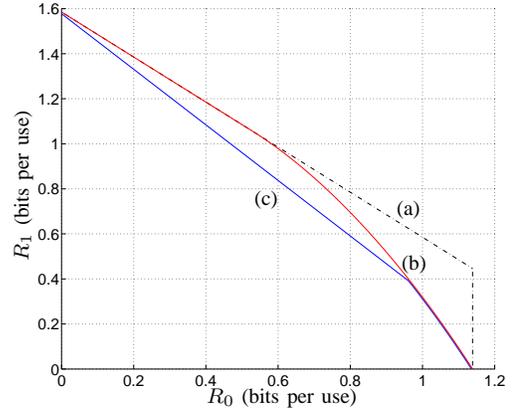, scale = 0.29}
\vspace{-0.2cm}
\caption{A single \textit{ternary} HD relay channel with two sources is considered. (a) Bound due to single source capacities. (b) Upper bound due to Theorem~\ref{th:C_two_sources}. (c) Region due to the coding strategy with block length $n = 640$.}
\vspace{-0.6cm}
\label{fig:rate_reg}
\end{figure}

\begin{remark}
Application of the \textit{binary} HD relay model yields $C = 0.5$ bits per channel use for all relay cascades composed of two or more \textit{binary} HD relays. Therefore, the optimum transmission strategy is just a straightforward time-sharing approach. The reason lies simply in the fact that the relays cannot encode parts of their information by means of the slot allocation since the subsequent relay is not able to recognize when nothing (i.~e. symbol N) was sent. 
\end{remark}
\begin{theorem}\label{th:C_two_sources}
The rate region of the relay network defined above with two sources, namely source node~$0$ and relay node~$r$, is characterized by
\begin{eqnarray}
\setcounter{equation}{5}
R_0 &\leq& \max_{p(\mathbf{x}_{[0:m]})}\min \left\{ H(Y_i|X_i): 1 \leq i \leq m\right\}\\
R_r &\leq& \max_{p(\mathbf{x}_{[0:m]})}\min \big\{H\left(Y_{i}|X_{r-1},X_{(i:i\leq m)}\right):\nonumber \\
&& \hspace{2.05cm}r+1 \leq i \leq m+1\big\}
\end{eqnarray}
and $(\ref{sum_rate_two_sources})$ shown at the bottom of the page. The maximization of the equations is performed jointly regarding $p(\mathbf{x}_{[0:m]})$.
\end{theorem}
\begin{proof}
The proof is given in the Appendix.
\end{proof}
\begin{example}[HD Single Relay Network with Two Sources]\label{example_3}
The \textit{ternary} relay network considered here is characterized by $m=1$ and $r=1$ and together with (\ref{relay_ch_model}), Theorem \ref{th:C_two_sources} becomes
\begin{eqnarray}
\setcounter{equation}{8}
R_0 &\leq& H(X_0|X_1) \\
R_1 &\leq&  H(X_1|X_0) \\
R_0 + R_1 &\leq& H(X_1).
\end{eqnarray}
An outer bound on the rate region of the considered line network is obviously given by $R_0 + R_1 \leq \textrm{log}_2 3$ bits (Fig.~\ref{fig:rate_reg}, graph~(a)) since the sum-rate can never be larger than the maximum of $H(X_1)$. We first try to determine whether points on this outer bound, besides $(R_0,R_1) = (0,\textrm{log}_2 3)$ bits, are delivered by equations (8) to (10) what inevitably requires a uniform $p_{X_1}(x_1)$. Since $H(X_0|X_1)$ has to be smaller or equal to $H(X_1)$, we are allowed to assume equality in (8) what follows from Theorem \ref{th:C_one_source}. By making the same optimality assumptions regarding $p(\mathbf{x}_{[0:1]})$ as in Example \ref{example_1}, we get $R_0 = \frac{1}{3}\textrm{log}_2 3$ bits and, consequently, $R_1 \leq \frac{2}{3}\textrm{log}_2 3$ bits. Note that this value for $R_1$ does not contradict with (9), i.~e. it is smaller than $H(X_1|X_0)$ concerning the assumed input distribution. The obtained point lies on the outer bound and it follows from a time-sharing argument that all points on the line between  $(0,\textrm{log}_2 3)$ bits and  $(\frac{1}{3}\textrm{log}_2 3,\frac{2}{3}\textrm{log}_2 3)$ bits are part of the rate region bound characterized by (8) to (10).

In the sequel, we maintain the optimality assumptions regarding $p(\mathbf{x}_{[0:1]})$ and focus on the remaining interval $\frac{1}{3}\textrm{log}_2 3 < R_0 \leq 1.1389$ bits, where $1.1389$ bits is the capacity of a single \textit{ternary} HD relay channel (Example \ref{example_1}). Again, $R_0 = H(X_0|X_1)$ but now $p_{X_1}(x_1)$ is not uniform anymore (due to $R_0>\frac{1}{3}\textrm{log}_2 3$) yielding a sum-rate strictly smaller than $\textrm{log}_2 3$ bits. An upper bound on $R_1$ is given by $H(X_1) - R_0$. It remains to check whether this expression is smaller or equal to the right hand side of (9) in the considered interval for $R_0$ for the assumed input distribution. However, this is satisfied and, therefore, the complete upper bound on the rate region according to (8)-(10) is characterized by
\begin{equation}
R_1 \leq \left\{\begin{array}{ll} \!\!\textrm{log}_2 3 - R_0, & \!\!\!0\leq R_0 \leq \frac{1}{3}\textrm{log}_2 3\\ \!\!H_b\left ( \frac{R_0}{\textrm{log}_2 3}\right) + \left(1- \frac{R_0}{\textrm{log}_2 3}\right) -R_0, & \!\!\!\frac{1}{3}\textrm{log}_2 3 < R_0 \leq C, \nonumber \end{array}\right.
\label{R_region_bound}
\end{equation}
where $H_b(\cdot)$ denotes the binary entropy function and $C$ = 1.1389 bits per channel use. A graphical representation is given in Fig. \ref{fig:rate_reg}, graph~(b).
\end{example}

\section{Coding Strategies}
\subsection{Achievability of $C$ in Theorem \ref{th:C_one_source}}

A coding strategy is presented capable of achieving $C$ in Theorem \ref{th:C_one_source}. As it is standard in achievability proofs, blocks of transmissions are used such that in $B$ blocks a sequence of $B-m$ indices $w_0 \in \mathcal{W}_0$ is sent from the source to the destination. As $B\rightarrow\infty$, the rate $\frac{R_0(B-m)}{B}\rightarrow R_0$. The idea behind the coding strategy is the following. Based on the feedforward property of the considered line network and due to the fact that each node is aware of the encoding strategy used by nodes with larger indices, node~$i$, $0 \leq i \leq m$, knows at each time instance the codeword, which will be sent by nodes $l>i$ in the upcoming transmission block. Thus, each node is able to adapt its transmission to the codeword chosen by the next node what can be exploited in order to prevent that concurrently sent codewords of adjacent nodes occupy the same time slots with binary symbols $\{0,1\}$. 

Different techniques for encoding are used by the source and the \textit{ternary} relays. While the source utilizes a ternary alphabet $\{0,1,\textrm{N}\}$ for encoding, the relays represent their messages by a combination of binary symbols $\{0,1\}$ and the allocation of binary symbols to the slots of a transmission block. Let $n_i$ denote the number of binary symbols used by relay $i$ during a single transmission block. Then, at most $2^{n_m} {n\choose n_m}$ indices can be encoded by relay $m$ where $2^{n_m}$ denotes the number of distinctive indices when the binary symbols are located at fixed slots while ${n \choose n_m}$ denotes the number of possible slot allocations. Due to the half-duplex constraint, the effective codeword length of relay $m-1$ reduces to $n-n_m$. This results from the fact that relay $m$ cannot pay attention to relay $m-1$ when relay $m$ sends binary symbols and, therefore, the number of indices, encodable by relay $m-1$, is at most $2^{n_{m-1}} {n -n_m\choose n_{m-1}}$. The same argumentation holds for each relay in the chain, i.~e. relay~$i$, $1 \leq i \leq m$, is able to encode at most $2^{n_{i}} {n -n_{i+1}\choose n_{i}}$ indices per transmission block where $n_{m+1} = 0$ since the sink node listens all the time. Finally, the effective length of the source codeword is $n-n_1$ what enables the source to encode a maximum of $3^{n-n_1}$ indices. The rate $R = n^{-1}\textrm{log}_2|\mathcal{W}_0|$ is
\begin{equation}
R_0 = \min \left \{ \frac{n-n_1}{n}\textrm{log}_2 3, \frac{n_i}{n} + \frac{1}{n}\textrm{log}_2{\,n -n_{i+1} \,\choose\, n_i}:\forall i \right\},
\label{rates_multi_relays}
\end{equation}
where $1 \leq i \leq m$.

\textit{Codebook Construction: }The source and all relays generate codewords according to the scheme described in the previous paragraph. Let $w_i \in \mathcal{W}_0$ indicate a message index forwarded by relay $i$, and let $s_i \in \mathcal{S}_i$ denote a particular slot allocation used by relay $i$ for encoding indices $w_i$. Note that each $s_i$ consists of $n-n_{i+1}$ slots, which can be embedded in at most $n \choose n_{i+1}$ ways into a block of length $n$ whereas the embedding is a function of the concurrently used $s_{i+1},\dots,s_m$. The resulting slot allocations of length $n$, employed by relay $i$, are denoted as $z_i \in \mathcal{Z}_i$ and depend on $s_i,\dots,s_m$. The procedure works as follows. Fix $|\mathcal{W}_0|$ relay $m$ codewords $x_m^n(w_m)$. For each slot allocation $z_m$ used in relay~$m$ codewords, construct $|\mathcal{W}_0|$ relay~$m-1$ codewords $x_{m-1}^n(w_{m-1},z_m)$. This ensures that relay $m-1$ can encode each message $w_{m-1}$ independently of the slot allocation used by relay $m$. The procedure repeats and, finally, for each slot allocation $z_1$ used in relay $1$ codewords, construct $|\mathcal{W}_0|$ source codewords $x_0^n(w_0,z_1)$.

\textit{Encoding (at the end of block $b-1$): } Let $w^{(b)}_0 \in \mathcal{W}_0$ denote the new message chosen by the source to be sent in block~$b$, and let $\hat{w}_i^{(b)} \in \mathcal{W}_0$ denote the estimate of $w^{(b-i)}_0$ made by relay~$i$ at the end of block $b-1$. Further, $\hat{s}_i^{(b)}$, which is a function of $\hat{w}_i^{(b)}$, corresponds to the slot allocation used by relay $i$ in transmission block $b$ for encoding $\hat{w}_i^{(b)}$ whereas $\hat{z}_i^{(b)}$ is determined by $\hat{s}_i^{(b)},\dots,\hat{s}_m^{(b)}$. Relay node $m$ sends $x_m^n(\hat{w}_m^{(b)})$ in block $b$. Since relay node $i$, $1 \leq i \leq m-1$, knows all previously sent indices $(\hat{w}_i^{(b-1)},\hat{w}_i^{(b-2)}\dots)$, which equal $(\hat{w}_{i+1}^{(b)},\hat{w}_{i+2}^{(b)},\dots)$, it knows $\hat{z}_{i+1}^{(b)}$ and encodes its latest index $\hat{w}_i^{(b)}$ with $x_{i}^n(\hat{w}_{i}^{(b)},\hat{z}_{i+1}^{(b)})$. Similarly, the source chooses $x_0^n(w^{(b)}_0,\hat{z}_1^{(b)})$ for transmission in block $b$.

\textit{Decoding (at the end of block $b-1$): } At the end of block~$b-2$, relay~$i$ has estimates $(\hat{w}_{i}^{(b-1)},\hat{w}_{i}^{(b-2)},\dots)$ and, therefore, estimates of $(\hat{s}_{i}^{(b-1)},\hat{s}_{i+1}^{(b-1)},\dots)$ and of $\hat{z}_i^{b-1}$. Then, based on the received sequence $x_{i-1}^n(\hat{w}_{i-1}^{(b-1)},\hat{z}_{i}^{(b-1)})$ during block~$b-1$ and due to the knowledge of the codebook used by relay~$i-1$, relay~$i$ is able to determine the unknown index $\hat{w}_{i-1}^{(b-1)}$. The destination knows the codebook used by relay $m$ and upon receiving $x_m^n(\hat{w}_m^{(b-1)})$, it can determine $\hat{w}_m^{(b-1)}$. Both the codebook construction and the noise freedom of the relay cascade guarantee, that the decoding steps can be performed with zero-error probability.

\begin{proof}[\hspace{-0.35cm}\textit{Achievability}] Using the relation $n^{-1}\textrm{log}{n \choose m} = H_b \left ( \frac{m}{n}\right)$ \cite[Th. 1.4.5]{lint99} as $n \rightarrow~\infty$, optimality assumptions regarding $p(\mathbf{x}_{[i:i+1]})$ (symmetry, zero probabilities - see Example \ref{example_2}), the resultant identities $\frac{n_i}{n} = p_{0\scriptsize{\textrm{N}}}^{i} + p_{1\scriptsize{\textrm{N}}}^{i}$ and $\frac{n-n_i-n_{i+1}}{n} = p_{\scriptsize{\textrm{N}}\scriptsize{\textrm{N}}}^{i}$, we obtain
\begin{eqnarray}
\frac{n_i}{n} + \frac{1}{n}\textrm{log}_2{n -n_{i+1}\choose n_i} &\longrightarrow& H\left(X_{i}|X_{(i+1:i+1\leq m)}\right), \nonumber
\end{eqnarray}
where $1\leq i \leq m$. According to the model in (\ref{relay_ch_model}), $H(X_i|X_{i+1}) = H(Y_{i+1}|X_{i+1})$ what shows that each entry in (\ref{rates_multi_relays}), except for the first, converges to the corresponding entry in (\ref{eq:C_one_source}). The first entry in (\ref{rates_multi_relays}) corresponds to a source, which uses uniformly distributed input symbols when relay~$1$ is listening. Evaluation of $H(Y_1|X_1)$ regarding a uniform $p_{X_0|X_1}(x_0,\textrm{N})$ yields $p_{X_1}(\textrm{N})\textrm{log}_23$. Hence, the first entry in (\ref{eq:C_one_source}) equals the first entry in (\ref{rates_multi_relays}).
\end{proof}
\begin{remark}
At this point, we are able to justify why it has been without loss of optimality to impose the Markov property on the channel inputs. Assume that each pair of channel inputs is statistically dependent given all remaining inputs. Then the procedure regarding Theorem \ref{th:C_one_source}, as shown in the Appendix, yields $\max \min \{H(Y_i|X_{[i:m]}), H(Y_{m+1}):1 \leq i \leq m\}$ as simplified cut-set bound what is smaller or at most equal to the achievable rate. But since the cut-set bound is an outer bound, only equality is valid, achieved e.~g. by $X_0\rightarrow\dots\rightarrow X_m$. For non-Markovian inputs, the rate region bound as stated in Theorem \ref{th:C_two_sources} is still an upper bound (but eventually looser). The Markov property merely cancels conditional random variables from the entropies what does not reduce the region.
\label{Remark_3}
\end{remark}
\subsection{Coding Strategy for a HD Relay Cascade with Two Sources}
A coding scheme based on similar ideas can be derived for a line network where a second relay node $r$ intends to transmit own information. Two main points have to be considered:
\begin{itemize}
\item Relay source $r$ and all subsequent relay nodes must be able to encode $|\mathcal{W}_0|\cdot |\mathcal{W}_r|$ different indices since $W_0$ and $W_r$ are independent.
\item The slot allocations $z_r \in \mathcal{Z}_r$, applied by relay source $r$, are completely determined by the source indices $w_0$.
\end{itemize}
\begin{theorem}
Consider a \textit{ternary} single HD relay channel where both source and relay send own information. The bound, described by equations (8) to (10), is achievable provided that the source rate exceeds a threshold.
\end{theorem}
\begin{proof}
Let $t\hspace{0.5mm}n_1$ and $(1-t)\hspace{0.5mm} n_1$ denote the number of binary symbols used by the relay for encoding each $w_0$ and $w_1$, respectively, where $0 \leq t \leq 1$. Further, all possible slot allocations of the relay represent indices $w_0$. If the number of source indices matches the number of relay codewords for representing source indices, or expressed in $R_0$
\begin{equation}
\frac{n-n_1}{n}\textrm{log}_23 = \frac{t\hspace{0.5mm}n_1}{n} + \frac{1}{n}\textrm{log}_2{n \choose n_1}, \quad 0 \leq t \leq 1,
\label{no_indices_equal}
\end{equation}
the cut-set bound is achievable. Note that the lhs of (\ref{no_indices_equal}) equals $p_{X_1}(\textrm{N})\textrm{log}_23$ what in turn equals $H(X_0|X_1)$, assumed the same $p(\mathbf{x}_{[0:1]})$ is used than in Example~\ref{example_1}. Further, $R_1 = (1-t)\hspace{0.5mm} n_1 n^{-1}$. As $n \rightarrow \infty$, $R_0 + R_1 \rightarrow H(X_1)$  what results from \cite[Th. 1.4.5]{lint99} under consideration of the rhs of (\ref{no_indices_equal}). The minimum $R_0$ (threshold) follows from (\ref{no_indices_equal}) for $t=0$.
\end{proof}
\section{Appendix}
\begin{proof}[Proof of Theorem \ref{th:C_one_source}]An upper bound on the capacity of each single source-destination network with source $0$ and sink node $m+1$ is given by \cite[Th. 14.10.1]{cover-thomas:it-book}
\begin{equation}
C \leq \max_{p(\mathbf{x}_{[0:m]})} \min_{S \in \mathcal{M}} I(X_0,X_{S^c};Y_{S},Y_{m+1}|X_{S}),
\label{cutset_bound_single_SD}
\end{equation}
where $\mathcal{M} = \mathcal{P}(\{1,\dots,m\})$ and $S^c$ is the complement of $S$ in $\{1,\dots,m\}$. In case of a noise-free network, (\ref{cutset_bound_single_SD}) becomes
\begin{equation}
C \leq \max_{p(\mathbf{x}_{[0:m]})}\min_{S \in \mathcal{M}} H(Y_{S},Y_{m+1}|X_{S}).
\label{App:cut-set_bound_exp}
\end{equation}
Let $S$ be nonempty and let $l \in \{1,\dots,m\}$ denote the smallest integer in $S$. Then
\begin{eqnarray}
 H(Y_{S},Y_{m+1}|X_{S}) \!\!\!&\stackrel{(a)}\geq&\!\!\! H(Y_l|X_{S_{\bar l}},X_l) + H(Y_{S_{\bar l}}|X_{S_{\bar l}},X_l,Y_l) \nonumber \\
\!\!\!&\stackrel{(b)}=&\!\!\! H(Y_l|X_l) + H(Y_{S_{\bar l}}|X_{S_{\bar l}},X_l,Y_l) \nonumber \\
\!\!\!&\stackrel{(c)}{\geq}&\!\!\! H(Y_l|X_l),
\label{App:cut-set_bound_simp}
\end{eqnarray}
where $(a)$ follows from the chain rule and $(b)$ from $X_{S_{\bar l}} \rightarrow X_l \rightarrow Y_l$. Equality in $(a)$ and $(c)$ is achieved by the ascending index sets $S = \{l,l+1,\dots,m\}$, $1 \leq l \leq m$, which compose the entries of a set say $\mathcal{M}_a$. Hence, for each $S^{\prime}\in \mathcal{M}\backslash\{\emptyset\}$ there exists an $S \in \mathcal{M}_a$ such that $H(Y_{S},Y_{m+1}|X_{S}) \leq H(Y_{S^\prime},Y_{m+1}|X_{S^\prime})$. Take e.~g. $S^\prime = \{l,l+v\}$, where $0 \leq v \leq m-l$, and extend it to an ascending index set $S = \{l,l+1,\dots,m\}$. The claim, stated in the sentence before the last, holds. In summary, (\ref{App:cut-set_bound_simp}) yields the first $m$ entries in (\ref{eq:C_one_source}) whereas the remaining entry, $H(Y_{m+1})$, follows when~$S$ in (\ref{App:cut-set_bound_exp}) is replaced by the empty set.
\end{proof}
\begin{proof}[Proof of Theorem \ref{th:C_two_sources}] 
The derivation of the individual rate bounds is almost along the same lines as in the proof of Theorem \ref{th:C_one_source}. Hence, we concentrate on the sum-rate bound. 

An upper bound on the sum-rate of each network with two sources $0$ and $r$ and a sink $m+1$ is \cite[Th. 14.10.1]{cover-thomas:it-book}
\begin{equation}
R_0 + R_r \leq \max_{p(\mathbf{x}_{[0:m]})}\min_{S \in \mathcal{M}} I(X_0,X_r,X_{S^c};Y_{S},Y_{m+1}|X_{S}), \\
\label{App_eq_sum_rate}
\end{equation}
where $\mathcal{M}$ is the power set of $M^d \cup M^u:=\{1,\dots,r-1\} \cup \{r+1,\dots,m\}$. Note that the rhs of (\ref{App_eq_sum_rate}) simplifies to the rhs of (\ref{App:cut-set_bound_exp}) due to the assumed noise freedom. Let $S^d \in \mathcal{P}(M^d)$ and $S^u \in \mathcal{P}(M^u)$ where $S = S^d \cup S^u$. First let $S^d$ and $S^u$ be nonempty, i.~e. $\mathcal{M}^\prime := \mathcal{P}(S) \subset \mathcal{M}$. Further, let $i$ and $j$ be the minimum and maximum values in $S^d$ whereas $k$ denotes the minimum value in $S^u$. Then
\begin{eqnarray}
\!\!\!\!\!\!\!\!\!H(Y_{S},Y_{m+1}|X_{S}) \!\!\!&\stackrel{(a)}\geq&\!\!\! H(Y_i,Y_k|X_S) \nonumber \\
\!\!\!&&\!\!\! \hspace{0.2cm}+  H(Y_{S^d_{\bar{i}}},Y_{S^u_{\bar{k}}}|X_S,Y_i,Y_k)\nonumber \\
\!\!\!&\stackrel{(b)}=&\!\!\! H(Y_i,Y_k|X_i,X_j,X_k)  \nonumber \\
\!\!\!&&\!\!\! \hspace{0.2cm}+ H(Y_{S^d_{\bar{i}}},Y_{S^u_{\bar{k}}}|X_S,Y_i,Y_k) \nonumber \\
\!\!\!&\stackrel{(c)}\geq&\!\!\! H(Y_i|X_i) + H(Y_k|X_{r-1},X_k),
\label{App_final_low_bound_two_sources}
\end{eqnarray}
where $(a)$ follows from the chain rule, $(b)$ from $(X_{S_{\bar i}^d},X_{S^u}) \rightarrow X_i \rightarrow Y_i$ and $(X_{S_{\bar j}^d},X_{S_{\bar k}^u}) \rightarrow (X_j,X_k) \rightarrow Y_k$, and $(c)$ from applying chain rule to the first term in $(b)$ under consideration of $(X_j,X_k) \rightarrow (X_{r-1},X_k) \rightarrow Y_k$ together with the described Markov relations. Equality in $(a)$ and $(c)$ is achieved by the ascending sets $S^d = \{i,i+1,\dots,r-1\}$, $1 \leq i \leq r-1$, and $S^u = \{k,k+1,\dots,m\}$, $r+1 \leq k \leq m$, which compose the entries $S = S^d \cup S^u$ of a set say $\mathcal{M}_a$. Then for each $S^{\prime}\in \mathcal{M}^\prime$ there exists a $S \in \mathcal{M}_a$ such that $H(Y_{S},Y_{m+1}|X_{S}) \leq H(Y_{S^\prime},Y_{m+1}|X_{S^\prime})$. Take e.~g. $S^\prime = \{i,i+v\}\cup\{k,k+w\}$, where $0 \leq v \leq r-1-i$ and $0 \leq w \leq m-k$, and extend $S^\prime$ to an ascending index set $S = \{i,i+1,\dots,r-1\}\cup\{k,k+1,\dots,m\}$. The inequality relation holds. In summary, the procedure yields 
\begin{eqnarray}
&&\min \{H(Y_i|X_i):1 \leq i \leq r-1\} \nonumber\\
&&\hspace{0.8cm}+ \min \{H(Y_k|X_{r-1},X_{k}):r+1 \leq k \leq m\} \nonumber
\end{eqnarray}
in (\ref{sum_rate_two_sources}), what follows from (\ref{App_final_low_bound_two_sources}) taking into account all  combinations of $i$ and $k$. The last entry in (\ref{sum_rate_two_sources}) and the modified version of above equation in (\ref{sum_rate_two_sources}) result when, in addition, the sets~$S \in \mathcal{M}\backslash \mathcal{M}^\prime$ are considered ($S^d$, $S^u$ empty or both).
\end{proof}

\bibliographystyle{unsrt}
\bibliography{bare_conf}

\end{document}